\documentclass[preprint,showpacs,aps,preprintnumbers,amsmath,amssymb]{revtex4-1}

\usepackage{graphicx,epsfig}% Include figure files
\usepackage{dcolumn}% Align table columns on decimal point
\usepackage{bm}% bold math
\newcommand{\be}{\begin{equation}}
\newcommand{\ee}{\end{equation}}
\newcommand{\bse}{\begin{subequations}}
\newcommand{\ese}{\end{subequations}}
\newcommand{\bary}{\begin{eqnarray}}
\newcommand{\eary}{\end{eqnarray}}
\newcommand{\bwt}{\begin{widetext}}
\newcommand{\ewt}{\end{widetext}}

\begin{document}

%\preprint{ICN/000-HEP}

\title{Photohadronic scenario in interpreting the February-March 2014 flare of 1ES 1011+496}
%{Very High Energy Cosmic Ray and Neutrino Events  from Centaurus A}
% Force line breaks with \\
\author{Sarira Sahu }
\email{sarira@nucleares.unam.mx}
\author{Alberto Rosales de Le\'{o}n}
\email{albertoros4@ciencias.unam.mx}
\author{Luis Salvador Miranda}
\email{luis.miranda@correo.nucleares.unam.mx}

\affiliation{
Instituto de Ciencias Nucleares, Universidad Nacional Aut\'onoma de M\'exico, 
Circuito Exterior, C.U., A. Postal 70-543, 04510 Mexico DF, Mexico\\
}
%\ead{sarira@nucleares.unam.mx}

\begin{abstract}
The extraordinary multi-TeV flare from 1ES 1011 +496 
during February-March 2014 was observed by MAGIC telescopes for 17
nights and the average spectrum of the whole period has a
non-trivial shape. We have
used the photohadronic model and a template EBL model to explain
the average spectrum which fits well to the flare data. The spectral
index $\alpha$ is the only free parameter in our model. 
We have also shown
that the non-trivial nature of the spectrum
is due to the change in the behavior of the optical depth
above $\sim 600$ GeV $\gamma$-ray energy accompanied with the high SSC
flux. 

\end{abstract}

%\pacs{98.70.Rz; 98.70.Sa; 98.70.Vc}% PACS, the Physics and Astronomy                     
\maketitle

\section{Introduction}

The 1ES 1011+496 (RA: $153.767^{\circ}$, DEC: $49.434^{\circ}$) is a high frequency peaked BL Lac (HBL) object at a
redshift of z= 0.212. This HBL was discovered at very high energy
(VHE)  $>$ 100 GeV by the MAGIC telescope in 2007 following an optical high state
reported by the Tuorla Blazar Monitoring Program\cite{Albert:2007kw}. Two
more multi-wavelength observations of the HBL were carried out by MAGIC
in 2008\cite{Ahnen:2016hsc} and in 2011-12\cite{Aleksic:2016wfj}. During these two observation periods the source did not show any
flux variability. On 5th February  2014, the VERITAS collaboration\cite{Weekes:2001pd} issued an alert 
about the flaring of 1ES 1011+496 which was immediately  followed
by MAGIC telescopes from February 6th to March 7th, a
total of 17 nights\cite{Ahnen:2016gog}. The flare was observed in the energy range
$\sim$ 75 GeV-3100 GeV and the flux could reach more than 10 times
higher than any previously  recorded flaring state of the source\cite{Albert:2007kw,Reinthal:2012gz}.
Despite this large variation, no significant intra-night variability
was observed in the flux. This allowed the collaboration to use the
average of the 17 nights observed spectral energy distribution (SED)
to look for the imprint of the extragalactic
background light (EBL) induced $\gamma$-rays absorption on it\cite{Ahnen:2016gog}. 

The light produced from all the sources in the universe throughout the
cosmic history pervades the intergalactic space which is now
at longer wavelengths due to the expansion of the Universe and
absorption/re-emission by dust and the light in the band 0.1--100
${\mu}$m 
is called the diffuse EBL\cite{Hauser:2001xs}.
The observed VHE spectrum of the distant sources are attenuated by
EBL producing $e^+e^-$ pairs. 
While the EBL is problematic for the study of high redshift VHE $\gamma$-ray
sources, at the same time the observed VHE $\gamma$-rays also provides an indirect method to
probe the EBL. The relation between the intrinsic VHE flux
$F_{\gamma, int}$ and the observed one $F_{\gamma, obs}$ are related through\cite{Hauser:2001xs,Dominguez:2010bv}
\be
F_{\gamma,obs}(E_{\gamma})=F_{\gamma,int}(E_{\gamma}) \,
e^{-\tau_{\gamma\gamma}(E_{\gamma},z)},
\label{fluxobsint}
\ee
where $\tau_{\gamma\gamma}$ is the optical depth. As the HBL 1ES 1011+496 is at a
 intermediate redshift, the observation of the VHE flare  from it will provide a good
 opportunity to study the EBL effect. Although a large number of different EBL
 models exist\cite{Stecker:1992wi,Salamon:1997ac,Franceschini:2008tp,Dominguez:2010bv,Dominguez:2013lfa},
 here we shall discuss two important models
 by Franceschini et al. \cite{Franceschini:2008tp} and Dominguez et al.\cite{Dominguez:2010bv,Dominguez:2013lfa}, which are used by Imaging
 Atmospheric Cherenkov Telescopes (IACTs) to study the EBL effect on
 the propagation of 
 high energy $\gamma$-rays.

The SEDs of the HBLs
have a double peak structure in the $\nu-\nu F_{\nu}$ plane.  While
the low  energy peak corresponds to
the synchrotron radiation from a population of relativistic electrons
in the jet, the high energy
peak believed to be due to the synchrotron self
Compton (SSC) scattering of the high energy electrons with their
self-produced synchrotron photons. 
The so called {\it
leptonic model} which incorporates both the synchrotron and SSC
processes in it is very successful in explaining the multi-wavelength emission from blazars and FR I
galaxies\cite{Fossati:1998zn,Ghisellini:1998it,Abdo:2010fk,Roustazadeh:2011zz,Dermer:1993cz,Sikora:1994zb}. However,
difficulties arise in explaining the
multi-TeV emission detected from many flaring AGN\cite{Aharonian:2009xn,Abramowski:2011ze,Krawczynski:2003fq,Cui:2004wi,Blazejowski:2005ih}
which shows that leptonic model 
may not be efficient in multi-TeV regime.

\section{Photohadronic Model}
We employ photohadronic model to explain the
multi-TeV flaring from  many HBLs\cite{Mucke:1998mk,Mucke:2000rn,Sahu:2013ixa,Sahu:2013cja,Sahu:2015tua}. Here
the standard interpretation of the leptonic model is used to explain the
low energy peaks. Thereafter, it is proposed that the
low energy tail of the SSC photons in the blazar jet
serve as the target for the Fermi-accelerated high energy protons,
within the jet to produce TeV photons through
the decay of $\pi^0$s from the $\Delta$-resonance\cite{Sahu:2013ixa}.
But the efficiency of the photohadronic process depends on
the photon density in the blazar jet. In a normal jet, the photon density is
low which makes the process inefficient. However, during the flaring, it is
assumed that the photon density in the inner jet region can go up so
that the $\Delta$-resonance production is moderately efficient. 
Here, the flaring occurs within a  compact and confined
volume of radius $R'_f$  (quantity with $^{\prime}$ implies in the jet
  comoving frame)  inside the blob of radius $R'_b$ ($R'_f <
R'_b$). The bulk Lorentz factor  in the inner jet should be larger
than the outer jet. But for simplicity we assume 
$\Gamma_{out}\simeq\Gamma_{in}\simeq \Gamma$. 
We cannot estimate the photon density in the inner jet region directly as it is
hidden. For simplicity, we assume the scaling behavior of the photon
densities in different background energies as follows\cite{Sahu:2013ixa,Sahu:2013cja,Sahu:2015tua}:
\be
{n'_{\gamma, f}(\epsilon'_{\gamma_1})}
{n'^{-1}_{\gamma, f}(\epsilon'_{\gamma_2})}
\simeq
{n'_\gamma(\epsilon'_{\gamma_1})}
{n'^{-1}_\gamma(\epsilon'_{\gamma_2})}.
\label{denscale}
\ee 
Above equation implies that the ratio of photon densities at two different
background energies $\epsilon'_{\gamma_1} $  and $\epsilon'_{\gamma_2} $ 
 in the flaring state ($n'_{\gamma, f}$) and in the non-flaring state
 ($n'_{\gamma}$) remain almost  the same. 
The photon density in the outer region is calculated from the observed
flux in the usual way. So the
unknown internal photon density is expressed in terms of the known photon density
calculated from the observed/fitted SED
in the SSC region which is again related to the observed flux in the same
region. This model explains very nicely the observed TeV flux from the orphan
 flares of 1ES1959+650, Markarian 421 as well as multi-TeV
 flaring from M87\cite{Sahu:2013ixa,Sahu:2013cja,Sahu:2015tua}.

In the observer frame, 
the $\pi^0$-decay photon energy
$E_{\gamma}$ and the 
background SSC photon energy $\epsilon_{\gamma}$ are
related through,
\be
E_\gamma \epsilon_\gamma \simeq 0.032\, {\cal D}^2\,(1+z)^{-2} ~{\rm
  GeV}^2,
\label{Eepsgamma}
\ee
where $E_{\gamma}$ satisfy the relation 
$E_p=10\Gamma{\cal D}^{-1} E_{\gamma}$.
${\cal D}\simeq \Gamma$ is the Doppler factor of the relativistic
jet  and  $E_p$ is the observed proton energy.
%(here we use natural units $c=\hbar=1$).
The intrinsic flux $F_{\gamma,int}$ of the flaring blazar is
proportional to a power-law with an exponential cut-off given as
$E^{-\alpha}_{\gamma}\, e^{-E_{\gamma}/E_{\gamma,c}}$, with the
spectral index $\alpha \ge 2$ and the cut-off energy is
$E_{\gamma,c}$\cite{Aharonian:2003be}.
The effect of both the exponential
cut-off and the EBL contribution are to reduce the VHE flux. For far-off
sources the EBL plays the dominant role 
which
shows that the $E_{\gamma,c}$ is much higher than the highest energy
$\gamma$-ray observed during the VHE flaring event.
Including EBL effect 
in the photohadronic
scenario\cite{Sahu:2015tua} the observed
multi-TeV  flux is expressed as 
\be
F_{\gamma,obs}(E_{\gamma})= A_{\gamma} \Phi_{SSC}(\epsilon_{\gamma}
)E^{-\alpha+3}_{\gamma,GeV} 
\,e^{-\tau_{\gamma\gamma}(E_{\gamma},z)}.
\label{sed}
\ee
The SSC energy
$\epsilon_{\gamma}$ and the observed energy $E_{\gamma}$ satisfy the condition given in
Eq. (\ref{Eepsgamma}),  $\Phi_{SSC}(\epsilon_{\gamma} )$ is the 
SSC flux corresponding to the energy $\epsilon_{\gamma}$ and
$E_{\gamma,GeV}$ implies $E_{\gamma}$ expressed in units of GeV and
$A_{\gamma}$ is the dimensionless normalization constant calculated
from the observed flare data\cite{Sahu:2015tua}.
The spectral index
$\alpha$ is the only free parameter here.
By comparing Eqs. (\ref{fluxobsint})
and (\ref{sed}) $F_{\gamma,int}$ can be obtained.

\begin{figure}%fig1
\vspace{-0.3cm}
{\centering
%\resizebox*{0.5\textwidth}{0.35\textheight}
\resizebox*{0.8\textwidth}{0.5\textheight}
{\includegraphics{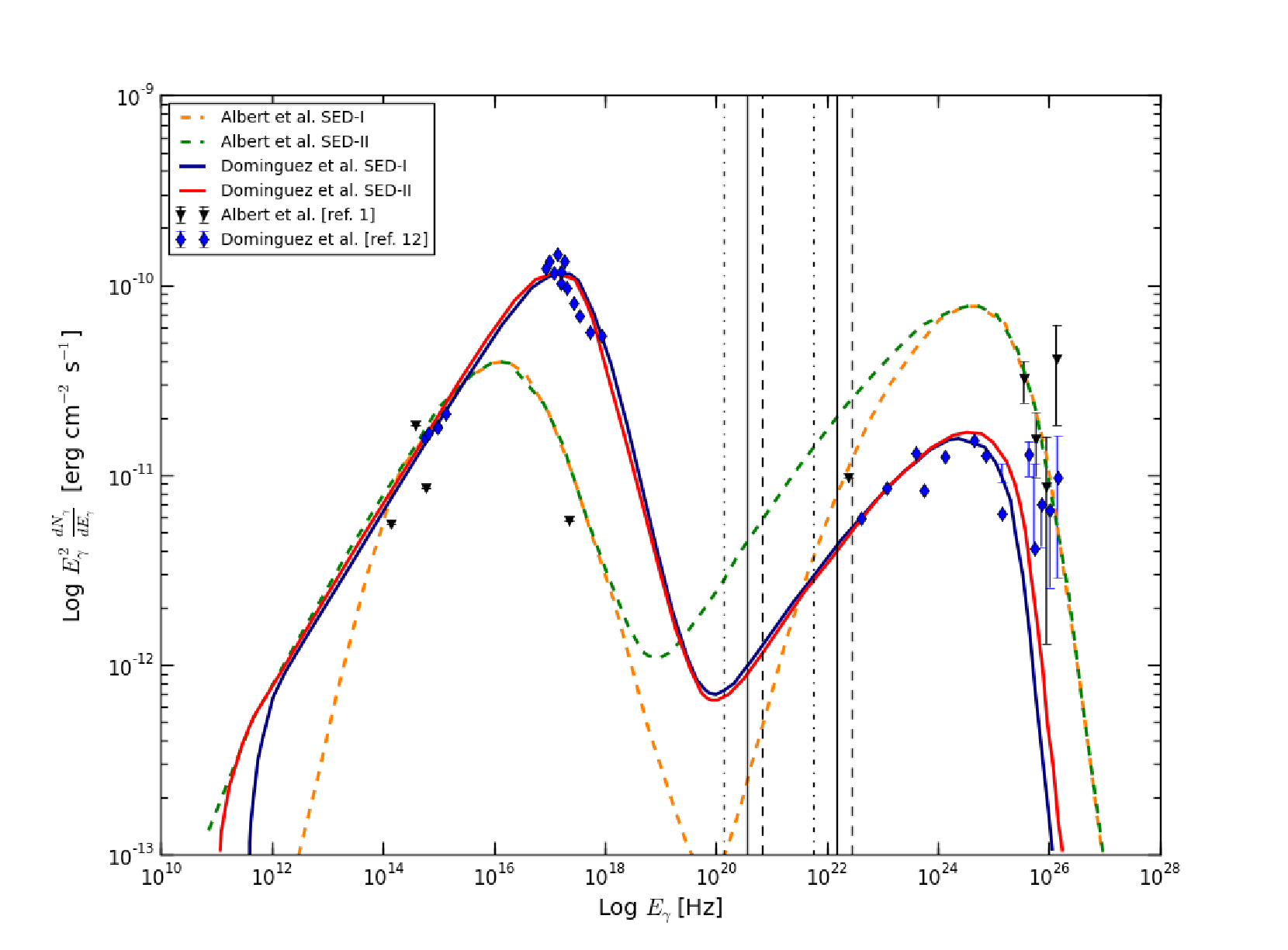}}
\par}
\caption{
The leptonic SED of the HBL 1ES 1011+496 is shown by using two
different models Albert et al. and  Dominguez et al.  Each of these models has two different parametrization
which we call as SED-I and SED-II. Also the regions in the 
SED where the VHE protons interact with the SSC photons to produce
the multi-TeV flare are shown: region between the two dashed dotted vertical
lines corresponds to SED-II of Dominguez et al. with ${\cal D}=9.1$, region
between the two vertical lines corresponds to SED-I of Dominguez et al. 
with ${\cal D}=14.6$
and the region between the two vertical dashed lines corresponds to
SED-I, II of  Albert et al. with ${\cal D}=20$.
\label{fig:supSED}
}
\end{figure}

\begin{figure}%fig2
\vspace{-0.3cm}
{\centering
%\resizebox*{0.5\textwidth}{0.35\textheight}
\resizebox*{0.8\textwidth}{0.5\textheight}
{\includegraphics{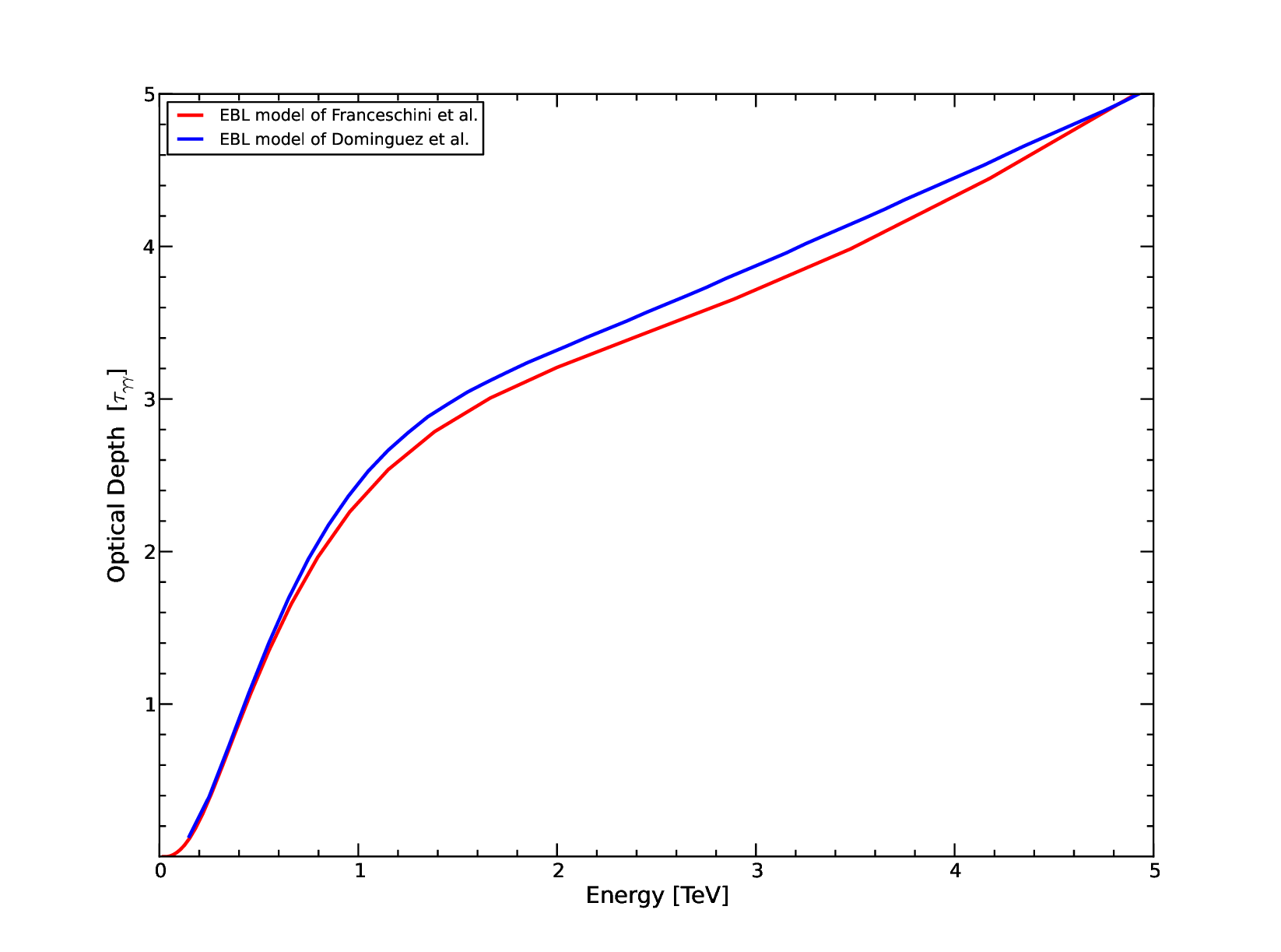}}
\par}
\caption{
At a redshift of $z=0.212$, the optical depth $\tau_{\gamma\gamma}$
in the EBL models of Dominguez et al. and Franceschini et al. are
shown for comparison.
\label{fig:supEBL}
}
\end{figure}

\section{Results}
The MAGIC collaboration fitted the  average of the 17 nights observed
SEDs of HBL 1ES 1011+496
with several functions, however, non of these fit well due to the non-trivial nature in the VHE
limit. Also the intrinsic SED is calculated
by subtracting the EBL contribution from the observed flux and is fitted with a simple power-law.
We use the photohadronic
scenario to interpret this flaring.
The input for the photohadronic process comes from the leptonic model
i.e. $\Gamma$, 
$\Phi_{SSC}$, and magnetic field  etc. We come across
two different leptonic models by Albert et
al.\cite{Albert:2007kw} and Dominguez et
al. \cite{Dominguez:2013lfa} which explain the low energy
SED of the HBL 1ES 1011+496 and each of them 
has two different
parametrization to fit the observed data as shown in Fig. \ref{fig:supSED}.
In Dominguez et al. model, 
the two different SEDs have almost the same flux in the SSC
energy range. So we only consider one of
the SEDs (SED-II) here.

The EBL models of Dominguez et al. and
Franceschini et al., are widely used to constraint the imprint
of EBL on the propagation of VHE
$\gamma$-rays by IACTs . We compared 
$\tau_{\gamma\gamma}$ of both these models (the central value of
the  former model is used) for $E_{\gamma} < 5$ TeV
and found a
very small difference as shown in
Fig. \ref{fig:supEBL}. So
for our analysis here we only consider the  Dominguez et al. model. However, the results will
be similar for the other one . There are three distinct regions of
$E_{\gamma}$ in Fig. \ref{fig:supEBL},
where the behavior of $\tau_{\gamma\gamma}$ is different. Below
$E_{\gamma}\sim 600$ GeV it has a rapid growth. In the energy range $\sim600$
GeV to $\sim 1.2$ TeV the growth is slow and above
$\sim 1.2$ TeV the growth is almost linear. This growth pattern of
$\tau_{\gamma\gamma}$ influences the $F_{\gamma,obs}$ in different
models and the  results of the above two leptonic models are discussed
separately.
%The combination of the EBL model
%and two different leptonic scenarios gives
%three different possibilities to fit the observed multi-TeV SED. 
%To simplify the interpretation, the results obtained in the  above two
%leptonic models are separately discussed.

\subsection{Leptonic model of Albert et al.\cite{Albert:2007kw} }
Here the SED is modeled  by using the single zone synchrotron-SSC
model where the emission region is a spherical blob of radius $R'_b\sim
10^{16}$ cm and a Doppler factor ${\cal D} = 20$ is taken. The emission
region has a magnetic field $B'\sim 0.15$ G and the relativistic
electrons emit synchrotron
radiation which explain the low energy peak of the SED. The high
energy emission from X-rays to few GeV $\gamma$-rays are from the
Compton scattering of the seed synchrotron photons by the same
population of high energy electrons. Here two
different SEDs are considered to fit the low energy data.
In the hadronic model alluded to previously, $ 75.6\, GeV \le E_{\gamma} \le 3.1\, TeV $
corresponds to the Fermi accelerated proton energy  in
the range $0.76\, TeV \le E_p \le 31\, TeV$ which collide with the SSC photons in the inner jet
region in the energy range
$115\, MeV\, (2.8\times 10^{22}\,Hz) \ge \epsilon_{\gamma} \ge 2.8\, MeV\, (6.8\times 10^{20}\,Hz)$ to produce the
$\Delta$-resonance and its decay to $\pi^0$s produces
observed multi-TeV $\gamma$-rays. Using the
scaling behavior of Eq. (\ref{denscale}),
photon densities in the inner and outer regions of the jet can be related.
In the outer region, the above range of $\epsilon_{\gamma}$ corresponds
to the low energy tail of the SSC photons 
(energy range between two dashed vertical lines in Fig. \ref{fig:supSED}). We 
observe that the $\Phi_{SSC}$ for SED-II is always
larger than the corresponding flux of SED-I. As we know from
Eq. (\ref{sed}), $F_{\gamma,obs}$  is proportional to $\Phi_{SSC}$, so
with the inclusion of  EBL contribution, 
the calculated $F_{\gamma,obs}$ with SED-II  is always $\ge$ the flux with SED-I
in the above range of $\epsilon_{\gamma}$. 

\begin{figure}%fig4
\vspace{-0.3cm}
{\centering
%\resizebox*{0.5\textwidth}{0.35\textheight}
\resizebox*{0.8\textwidth}{0.5\textheight}
{\includegraphics{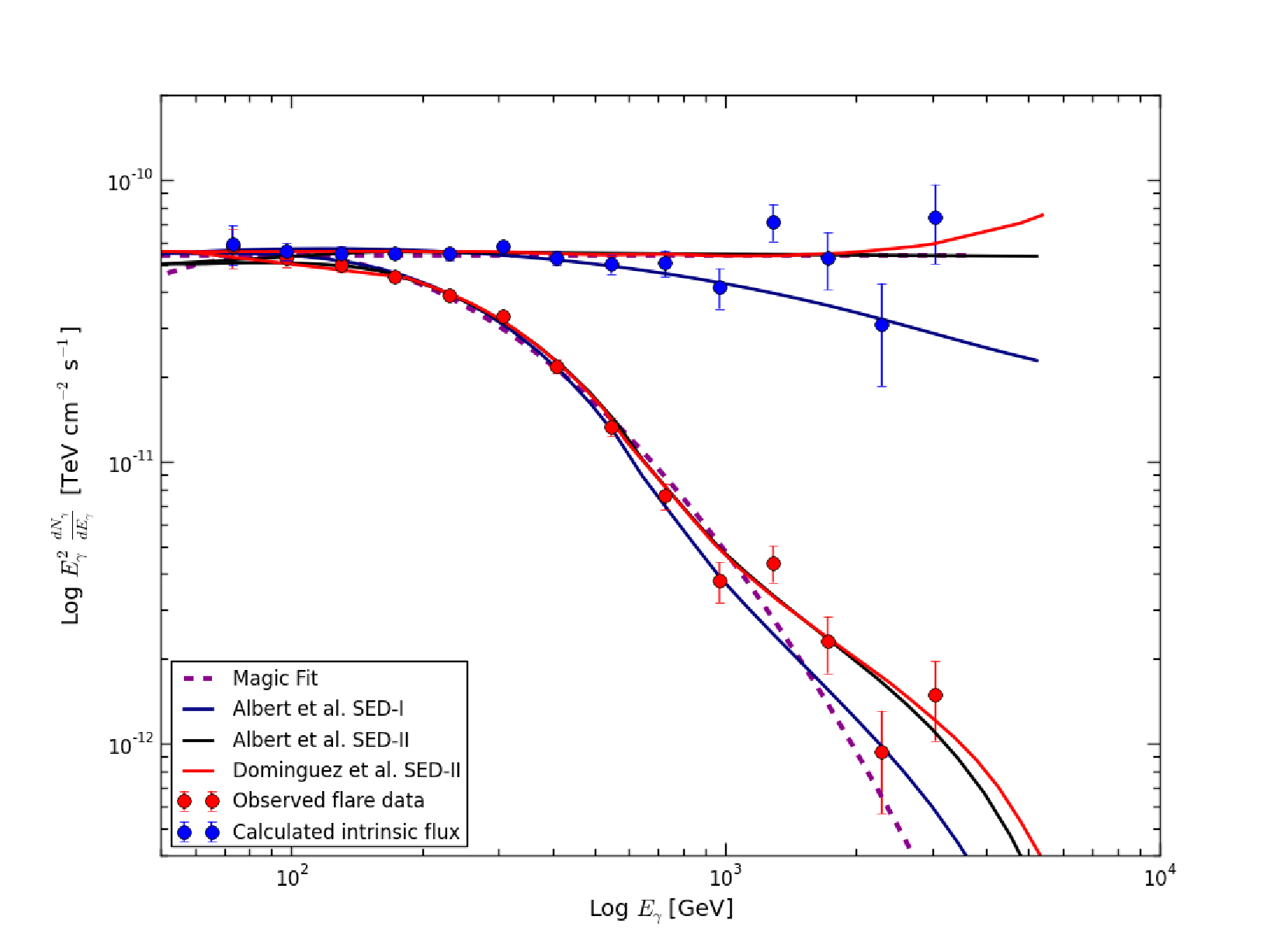}}
\par}
\caption{
Using leptonic models of Albert et al. and Dominguez et al. and the EBL
correction,  the multi-TeV flare data are fitted in the photohadronic model
(lower curves) and the corresponding intrinsic fluxes are also shown
(upper curves). The lower and the upper curves of same color belong to
a single model. For comparison
the MAGIC fit to the observed flux (lower magenta dashed curve) and the
intrinsic flux (upper magenta dashed  curve) are shown.
\label{fig:VHEAlb1} 
}
\end{figure}

The $F_{\gamma,obs}$ and $F_{\gamma,int}$ for SED-I are plotted as
functions of $E_{\gamma}$ in Fig. \ref{fig:VHEAlb1}. A good fit to
flare data is obtained for the normalization constant
$A_{\gamma}=0.37$ and the spectral index $\alpha=2.3$
(blue curves).
Our model fits very well with the flare data up to energy
$E_{\gamma} \sim 1$ TeV and above this energy the
flux falls faster than the observed data. 
Above 500 GeV the $F_{\gamma,int}$ (upper blue curve) falls faster than the MAGIC
fit which is a constant. This fall in $F_{\gamma,int}$ is also responsible for the
faster fall in $F_{\gamma,obs}$ in the energy range $\sim 500$ GeV
to 1.2 TeV even if the fall in $e^{-\tau_{\gamma\gamma}}$ is
slow. Above $E_{\gamma}\sim 1.2$ TeV, the linear growth in
$\tau_{\gamma\gamma}$ wins over the fall in  $F_{\gamma,int}$ so that
the fall in  $F_{\gamma,obs}$ is slowed down. 
For comparison we have also shown the log-parabola fit by MAGIC
collaboration (lower magenta dashed curve), however, both
these fits are poor in VHE limit.

We have also plotted $F_{\gamma,obs}$ and $F_{\gamma,int}$ for SED-II. Here a good fit is obtained for
$A_{\gamma}=0.64$ and $\alpha=2.6$ (lower black curve). We observed that 
the MAGIC fit to $F_{\gamma,int}$ and our result (upper black curve) are the
same and constant in all the energy range. In the photohadronic model,
above $\sim 1$ TeV the $F_{\gamma,obs}$ has a slow fall even though the
$F_{\gamma,int}$ is constant for all energies. Again the
curve changes its behavior above $\sim 1.2$ TeV. This peculiar behavior 
is due the slow growth of $\tau_{\gamma\gamma}$ in the range
$600\, GeV \le E_{\gamma}\sim 1.2$ TeV and above this energy almost a linear growth.
The comparison of $F_{\gamma,obs}$  in SED-I and SED-II shows a
marked difference for $E_{\gamma} > 0.8$ TeV. The lower black curve (SED-II) falls
slower than the lower blue curve (SED-I) . 
The higher value of $\Phi_{SSC}$ in SED-II compared to the
one in SED-I in the energy range $115\, MeV \ge \epsilon_{\gamma} \ge 2.8\, MeV$ is responsible for
this discrepancy which can be seen from Fig.  \ref{fig:supSED}.

\subsection{Leptonic model of Dominguez et
  al.\cite{Dominguez:2013lfa}}
As discussed above, this model uses two different parameterizations to fit the 
leptonic SED which we call as SED-I and SED-II as shown in
Fig. \ref{fig:supSED}. The SEDs obtained in
both these cases are almost the same in the SSC energy range. So here we only consider SED-II. 
However, for SED-I the results will be very similar.
The SED-II is fitted by considering the spherical blob of size
$R'_b=2.2\times 10^{16}$ cm moving with a bulk Lorentz factor
$\Gamma=9.1$. A constant magnetic field $B' \sim 0.23$ G is 
present in the blob region where the charged particles undergo
synchrotron emission.

In the photohadronic scenario, the flare 
energy range $75.6\, GeV \le E_{\gamma} \le 3.1\, TeV$
corresponds to 
$23.9\, MeV\, (5.8\times 10^{21}\,Hz) \ge \epsilon_{\gamma} \ge 0.58\,
MeV\, (1.4\times 10^{20}\,Hz)$ and the VHE proton energy
in the range $0.76\, TeV \le E_p \le 31\, TeV$. The above range of
$\epsilon_{\gamma}$ lies
in the tail region of the SSC spectrum as shown in
Fig. \ref{fig:supSED}.
In Fig. \ref{fig:VHEAlb1} we have also shown $F_{\gamma,obs}$ and $F_{\gamma,int}$ for SED-II.
A good fit to flare data is obtained by taking $A_{\gamma}=5.9$ and
$\alpha=2.6$ (lower red curve). We observed that our model fit decreases
slower than the MAGIC fit and model fits of Albert et al. above $\sim
1$ TeV. The comparison of $F_{\gamma,int}$ (upper red curve) with the MAGIC
fit shows that both are practically the same for $E_{\gamma} < 2$ TeV and above this energy
the photohadronic prediction increases slightly, however, there is a
big difference in $F_{\gamma,obs}$ above $E_{\gamma} > 1$ TeV.
From Eq. (\ref{sed}) we observed that both the intrinsic and the
observed fluxes are proportional to $E^{-\alpha}_{\gamma}$ and are
independent of an exponential cut-off. However, if at all there is a
cut-off energy it must be $E_{\gamma,c}\ge 70$ TeV, otherwise the 
$F_{\gamma,obs}$ will fall faster than the predicted fluxes shown in
black and red lower curves in Fig. \ref{fig:VHEAlb1} which will be non compatible with the flare data.

\section{Conclusions}
The multi-TeV flaring of
February-March 2014 from 1ES 1011+496 is interpreted using the photohadronic
scenario. To account for the effect of the diffuse radiation background on the VHE
$\gamma$-rays we incorporate a template EBL model to
calculate the observed flux. 
%This scenario depends on the SSC flux from the
%leptonic process and the spectral index $\alpha$ is the only free
%parameter. 
Also two different leptonic models are considered to
fit the flare data and the results are compared. The spectral index
$\alpha$ is the only free parameter here.
The flare data has a non-trivial shape above $E_{\gamma}\sim
600$ GeV and in photohadronic model this behavior can be explained by
the slow to linear growth in $\tau_{\gamma\gamma}$ above this energy
range complemented by higher SSC flux.
The EBL contribution alone cannot explain the non-trivial shape of the
data which can be clearly seen by comparing the lower blue curve with
the lower black and red curves in Fig. \ref{fig:VHEAlb1}.
Towards the end of the observation period by the MAGIC telescopes,
the source activity was lower which amounted to larger uncertainties in
the flux and correspondingly the average spectrum. Probably
this might be the reason for larger uncertainties in the VHE range of the
average spectrum. The MAGIC telescopes exposure period for most of the nights was $\sim
40$ minutes which was extended for $\sim 2$ hours on nights of 8th
and 9th February\cite{Ahnen:2016gog}. This extended period of observation might have
better flux resolution and our expectation is that photohadronic scenario will be able to
fit the data well. In future, for a better understanding of the EBL
effect and the role played by the SSC photons on the VHE
$\gamma$-ray flux from intermediate to high redshift blazars, it is necessary
to have simultaneous observations in multi-wavelength to the flaring objects. 

We thank Adiv Gonzalez and Lucy Fortson for many useful
discussions. The work of S. S. is partially supported by
DGAPA-UNAM (Mexico) Project No. IN110815.


\begin{thebibliography}{55}
\expandafter\ifx\csname natexlab\endcsname\relax\def\natexlab#1{#1}\fi
\expandafter\ifx\csname bibnamefont\endcsname\relax
  \def\bibnamefont#1{#1}\fi
\expandafter\ifx\csname bibfnamefont\endcsname\relax
  \def\bibfnamefont#1{#1}\fi
\expandafter\ifx\csname citenamefont\endcsname\relax
  \def\citenamefont#1{#1}\fi
\expandafter\ifx\csname url\endcsname\relax
  \def\url#1{\texttt{#1}}\fi
\expandafter\ifx\csname urlprefix\endcsname\relax\def\urlprefix{URL }\fi
\providecommand{\bibinfo}[2]{#2}
\providecommand{\eprint}[2][]{\url{#2}}

%1
%\cite{Albert:2007kw}
\bibitem{Albert:2007kw} 
  J.~Albert {\it et al.} [MAGIC Collaboration],
  %``Discovery of Very High Energy gamma-rays from 1ES1011+496 at z=0.212,''
  Astrophys.\ J.\  {\bf 667}, L21 (2007).
%  doi:10.1086/521982
%  [arXiv:0706.4435 [astro-ph]].
  %%CITATION = doi:10.1086/521982;%%
  %89 citations counted in INSPIRE as of 02 Mar 2016
%2
%\cite{Ahnen:2016hsc}
\bibitem{Ahnen:2016hsc} 
  M.~L.~Ahnen {\it et al.} [MAGIC and AGILE Collaborations],
  %``Multi-Wavelength Observations of the Blazar 1ES 1011+496 in Spring 2008,''
  Mon.\ Not.\ Roy.\ Astron.\ Soc.\  {\bf 459}, 2286 (2016).
%  doi:10.1093/mnras/stw710
 % [arXiv:1603.07308 [astro-ph.HE]].
  %%CITATION = doi:10.1093/mnras/stw710;%%
  %2 citations counted in INSPIRE as of 12 Sep 2016
% 3 PROBLEMA
%\cite{Aleksic:2016wfj}
\bibitem{Aleksic:2016wfj} 
 J.~ Aleksi{\'c}, {\it et al.}, Astron.\ Astrophys. {\bf 591,} A10 (2016). 
% Insights into the emission of the blazar 1ES
%1011+496 through unprecedented broadband observations during 2011 and
%2012.
%4
%\cite{Weekes:2001pd}
\bibitem{Weekes:2001pd} 
  T.~C.~Weekes {\it et al.},
  %``VERITAS: The Very energetic radiation imaging telescope array system,''
  Astropart.\ Phys.\  {\bf 17}, 221 (2002).
%  doi:10.1016/S0927-6505(01)00152-9

 % [astro-ph/0108478].
  %%CITATION = doi:10.1016/S0927-6505(01)00152-9;%%
  %229 citations counted in INSPIRE as of 13 Sep 2016
%5
%\cite{Ahnen:2016gog}
\bibitem{Ahnen:2016gog} 
  M.~L.~Ahnen {\it et al.},
  %``MAGIC observations of the February 2014 flare of 1ES 1011+496 and ensuing constraint of the EBL density,''
  Astron.\ Astrophys.\  {\bf 590}, A24 (2016).
%  doi:10.1051/0004-6361/201527256
 % [arXiv:1602.05239 [astro-ph.HE]].
  %%CITATION = doi:10.1051/0004-6361/201527256;%%
  %5 citations counted in INSPIRE as of 13 Sep 2016
%6  
%\cite{Reinthal:2012gz}
\bibitem{Reinthal:2012gz} 
  R.~Reinthal {\it et al.} [MAGIC and AGILE Team Collaborations],
  %``Multi-Wavelength Observations of the HBL Object 1ES 1011+496 in Spring 2008,''
  J.\ Phys.\ Conf.\ Ser.\  {\bf 355}, 012017 (2012).
 % doi:10.1088/1742-6596/355/1/012017
  %[arXiv:1109.6504 [astro-ph.HE]].
  %%CITATION = doi:10.1088/1742-6596/355/1/012017;%%
  %10 citations counted in INSPIRE as of 13 Sep 2016
%7
%\cite{Hauser:2001xs}
\bibitem{Hauser:2001xs} 
  M.~G.~Hauser and E.~Dwek,
  %``The cosmic infrared background: measurements and implications,''
  Ann.\ Rev.\ Astron.\ Astrophys.\  {\bf 39}, 249 (2001).
 % doi:10.1146/annurev.astro.39.1.249.
  %[astro-ph/0105539].
  %%CITATION = doi:10.1146/annurev.astro.39.1.249;%%
  %418 citations counted in INSPIRE as of 13 Sep 2016
% 8
%\cite{Dominguez:2010bv}
\bibitem{Dominguez:2010bv} 
  A.~Dominguez {\it et al.},
  %``Extragalactic Background Light Inferred from AEGIS Galaxy SED-type Fractions,''
  Mon.\ Not.\ Roy.\ Astron.\ Soc.\  {\bf 410}, 2556 (2011).
 % doi:10.1111/j.1365-2966.2010.17631.x
  %[arXiv:1007.1459 [astro-ph.CO]].
  %%CITATION = doi:10.1111/j.1365-2966.2010.17631.x;%%
  %233 citations counted in INSPIRE as of 13 Sep 2016
%9
%\cite{Salamon:1997ac}
\bibitem{Salamon:1997ac} 
  M.~H.~Salamon and F.~W.~Stecker,
  %``Absorption of high-energy gamma-rays by interactions with starlight photons in extragalactic space at high redshifts and the high-energy gamma-ray background,''
  Astrophys.\ J.\  {\bf 493}, 547 (1998).
%  doi:10.1086/305134
% & [astro-ph/9704166].
  %%CITATION = doi:10.1086/305134;%%
  %168 citations counted in INSPIRE as of 30 Sep 2016

%10
%\cite{Stecker:1992wi}
\bibitem{Stecker:1992wi} 
  F.~W.~Stecker, O.~C.~de Jager and M.~H.~Salamon,
  %``TeV gamma rays from 3C 279 - A possible probe of origin and intergalactic infrared radiation fields,''
  Astrophys.\ J.\  {\bf 390}, L49 (1992).
%  doi:10.1086/186369
  %%CITATION = doi:10.1086/186369;%%
  %315 citations counted in INSPIRE as of 30 Sep 2016
%11
%\cite{Franceschini:2008tp}
\bibitem{Franceschini:2008tp} 
  A.~Franceschini, G.~Rodighiero and M.~Vaccari,
  %``The extragalactic optical-infrared background radiations, their time evolution and the cosmic photon-photon opacity,''
  Astron.\ Astrophys.\  {\bf 487}, 837 (2008).
%  doi:10.1051/0004-6361:200809691
 % [arXiv:0805.1841 [astro-ph]].
  %%CITATION = doi:10.1051/0004-6361:200809691;%%
  %451 citations counted in INSPIRE as of 13 Sep 2016
%12
%\cite{Dominguez:2013lfa}
\bibitem{Dominguez:2013lfa} 
  A.~Dominguez, J.~D.~Finke, F.~Prada, J.~R.~Primack, F.~S.~Kitaura, B.~Siana and D.~Paneque,
  %``Detection of the cosmic \gamma-ray horizon from multiwavelength observations of blazars,''
  Astrophys.\ J.\  {\bf 770}, 77 (2013).
%  doi:10.1088/0004-637X/770/1/77
 % [arXiv:1305.2162 [atro-ph.CO]].
  %%CITATION = doi:10.1088/0004-637X/770/1/77;%%
  %38 citations counted in INSPIRE as of 13 Sep 2016
%13
%\cite{Abdo:2010fk}
\bibitem{Abdo:2010fk}
  A.~A.~Abdo {\it et al.}  [Fermi LAT Collaboration],
  %``Fermi Large Area Telescope View of the Core of the Radio Galaxy Centaurus A,''
  Astrophys.\ J.\  {\bf 719}, 1433-1444 (2010).
 % [arXiv:1006.5463 [astro-ph.HE]].
%14
%\cite{Roustazadeh:2011zz}
\bibitem{Roustazadeh:2011zz}
  P.~Roustazadeh and M.~B\"ottcher,
  %``Very high energy gamma-ray-induced pair cascades in the radiation fields of
  %dust Tori of active galactic nuclei: Application to Cen A,''
  Astrophys.\ J.\  {\bf 728}, 134 (2011).
  %%CITATION = ASJOA,728,134;%%
%15
%\cite{Fossati:1998zn}
\bibitem{Fossati:1998zn}
  G.~Fossati, L.~Maraschi, A.~Celotti, A.~Comastri and G.~Ghisellini,
  %``A Unifying view of the spectral energy distributions of blazars,''
  Mon.\ Not.\ Roy.\ Astron.\ Soc.\  {\bf 299} (1998) 433.
 % [arXiv:astro-ph/9804103].
  %%CITATION = MNRAA,299,433;%%
%16
%\cite{Ghisellini:1998it}
\bibitem{Ghisellini:1998it}
  G.~Ghisellini, A.~Celotti, G.~Fossati, L.~Maraschi and A.~Comastri,
  %``A Theoretical unifying scheme for gamma-ray bright blazars,''
  Mon.\ Not.\ Roy.\ Astron.\ Soc.\  {\bf 301} (1998) 451.
%  [arXiv:astro-ph/9807317].
  %%CITATION = MNRAA,301,451;%%
%17
%\cite{Dermer:1993cz}
\bibitem{Dermer:1993cz} 
  C.~D.~Dermer and R.~Schlickeiser,
  %``Model for the high-energy emission from blazars,''
  Astrophys.\ J.\  {\bf 416}, 458 (1993).
  %%CITATION = ASJOA,416,458;%%
  %334 citations counted in INSPIRE as of 30 Dec 2014
%18
%\cite{Sikora:1994zb}
\bibitem{Sikora:1994zb} 
  M.~Sikora, M.~C.~Begelman and M.~J.~Rees,
  %``Comptonization of diffuse ambient radiation by a relativistic jet: The source of gamma rays from blazars?,''
  Astrophys.\ J.\  {\bf 421}, 153 (1994).
  %%CITATION = ASJOA,421,153;%%
  %516 citations counted in INSPIRE as of 30 Dec 2014
%19
%\cite{Aharonian:2009xn}
\bibitem{Aharonian:2009xn} 
  F.~Aharonian {\it et al.}  [HESS Collaboration],
  %``Discovery of very high energy gamma-ray emission from Centaurus A with H.E.S.S,''
  Astrophys.\ J.\  {\bf 695}, L40 (2009).
%  [arXiv:0903.1582 [astro-ph.CO]].
  %%CITATION = ARXIV:0903.1582;%%
  %127 citations counted in INSPIRE as of 23 Dec 201
%20
%\cite{Abramowski:2011ze}
\bibitem{Abramowski:2011ze} 
  A.~Abramowski {\it et al.}  [H.E.S.S. and VERITAS Collaborations],
  %``The 2010 very high energy gamma-ray flare \& 10 years of multi-wavelength observations of M 87,''
  Astrophys.\ J.\  {\bf 746}, 151 (2012).
  %[arXiv:1111.5341 [astro-ph.CO]].
  %%CITATION = ARXIV:1111.5341;%%
  %29 citations counted in INSPIRE as of 01 Oct 201
%21
%\cite{Krawczynski:2003fq}
\bibitem{Krawczynski:2003fq} 
  H.~Krawczynski, S.~B.~Hughes, D.~Horan, F.~Aharonian, M.~F.~Aller, H.~Aller, P.~Boltwood and J.~Buckley {\it et al.},
  %``Multiwavelength observations of strong flares from the TeV - blazar 1ES 1959+650,''
  Astrophys.\ J.\  {\bf 601}, 151 (2004).
%  [astro-ph/0310158].
  %%CITATION = ASTRO-PH/0310158;%%
  %169 citations counted in INSPIRE as of 23 Dec 2014
%22
%\cite{Cui:2004wi}
\bibitem{Cui:2004wi} 
  W.~Cui {\it et al.}  [VERITAS Collaboration],
  %``News from a multi-wavelength monitoring campaign on Mrk 421,''
  AIP Conf.\ Proc.\  {\bf 745}, 455 (2005).
%  [astro-ph/0410160].
  %%CITATION = ASTRO-PH/0410160;%%
  %5 citations counted in INSPIRE as of 30 Dec 2014  
%23
%\cite{Blazejowski:2005ih}
\bibitem{Blazejowski:2005ih} 
  M.~Blazejowski, G.~Blaylock, I.~H.~Bond, S.~M.~Bradbury, J.~H.~Buckley, D.~A.~Carter-Lewis, O.~Celik and P.~Cogan {\it et al.},
  %``A Multi-wavelength view of the TeV blazar Markarian 421: Correlated variability, flaring, and spectral evolution,''
  Astrophys.\ J.\  {\bf 630}, 130 (2005).
 % [astro-ph/0505325].
  %%CITATION = ASTRO-PH/0505325;%%
  %94 citations counted in INSPIRE as of 23 Dec 2014
% 24
%\cite{Mucke:1998mk}
\bibitem{Mucke:1998mk} 
  A.~Mucke, J.~P.~Rachen, R.~Engel, R.~J.~Protheroe and T.~Stanev,
  %``On photohadronic processes in astrophysical environments,''
  Publ.\ Astron.\ Soc.\ Austral.\  {\bf 16}, 160 (1999).
  %doi:10.1071/AS99160
  %[astro-ph/9808279].
  %%CITATION = doi:10.1071/AS99160;%%
  %45 citations counted in INSPIRE as of 19 Jan 2016
%25
%\cite{Mucke:2000rn}
\bibitem{Mucke:2000rn} 
  A.~Mucke and R.~J.~Protheroe,
  %``A Proton synchrotron blazar model for flaring in Markarian 501,''
  Astropart.\ Phys.\  {\bf 15}, 121 (2001).
 % doi:10.1016/S0927-6505(00)00141-9
  %[astro-ph/0004052].
  %%CITATION = doi:10.1016/S0927-6505(00)00141-9;%%
  %181 citations counted in INSPIRE as of 18 Jan 2016
%26
%\cite{Sahu:2013ixa}
\bibitem{Sahu:2013ixa} 
  S.~Sahu, A.~F.~O.~Oliveros and J.~C.~Sanabria,
  %``Hadronic-origin orphan TeV flare from 1ES 1959+650,''
  Phys.\ Rev.\ D {\bf 87}, 103015 (2013).
 % [arXiv:1305.4985 [hep-ph]].
  %%CITATION = ARXIV:1305.4985;%%
  %6 citations counted in INSPIRE as of 23 Dec 2014
%27
%\cite{Sahu:2015tua}
\bibitem{Sahu:2015tua} 
  S.~Sahu, L.~S.~Miranda and S.~Rajpoot,
  %``Multi-TeV flaring from blazars: Markarian 421 as a case study,''
  Eur.\ Phys.\ J.\ C {\bf 76}, 127 (2016).
%  doi:10.1140/epjc/s10052-016-3975-2
 % [arXiv:1501.00973 [astro-ph.HE]].
  %%CITATION = doi:10.1140/epjc/s10052-016-3975-2;%%
  %1 citations counted in INSPIRE as of 13 Sep 2016
%28
%\cite{Sahu:2013cja}
\bibitem{Sahu:2013cja} 
  S.~Sahu and E.~Palacios,
  %``Hadronic origin of the TeV flare of M87 in April 2010,''
  Eur.\ Phys.\ J.\ C {\bf 75}, 52 (2015).
 % [arXiv:1310.1381 [astro-ph.HE]].
  %%CITATION = ARXIV:1310.1381;%%
  %2 citations counted in INSPIRE as of 13 Feb 2015
%29
%\cite{Aharonian:2003be}
\bibitem{Aharonian:2003be} 
  F.~Aharonian {\it et al.}  [HEGRA Collaboration],
  %``Detection of TeV gamma-rays from the bl lac 1es1959+650 in its low states and during a major outburst in 2002,''
  Astron.\ Astrophys.\  {\bf 406}, L9 (2003).
%  [astro-ph/0305275].

\end{thebibliography}
\end{document}